\documentclass[aps,prl,twocolumn,bibnotes,superscriptaddress,tightenlines,groupedaddress,showpacs]{revtex4}
\usepackage{amsmath,amsfonts,amssymb,amsbsy,graphicx,epsfig,bbm,times,hyperref}
\newcommand{\eq}[1]{Eq.~(\ref{#1})}

\newcommand{\ro}{\hat{\varrho}}

\newcommand{\gr}[1]{\boldsymbol{#1}}

\newcommand{\be}{\begin{equation}}
\newcommand{\ee}{\end{equation}}

\newcommand{\tr}{{\rm Tr}}

\newcommand{\ket}[1]{\left|{#1}\right\rangle}
\newcommand{\bra}[1]{\left\langle{#1}\right|}

\newcommand{\avr}[1]{\langle#1\rangle}

\begin{document}

\title{Optimal Quantum Estimation of the Unruh-Hawking Effect}
\author{Mariona Aspachs}
\affiliation{Grup de F\'isica Te\`orica, Universitat Aut\`onoma de Barcelona, 08193 Bellaterra (Barcelona), Spain}
\author{Gerardo Adesso}\author{Ivette Fuentes}\thanks{Previously known as Fuentes-Guridi and Fuentes-Schuller.}
\affiliation{{School of Mathematical Sciences, University of Nottingham, University Park, Nottingham NG7 2RD, United Kingdom}}

\date{October 8, 2010}

\begin{abstract}
We address on general quantum-statistical grounds the problem of optimal detection of the Unruh--Hawking effect. We show that the effect signatures are magnified up to potentially observable levels if the scalar field to be probed has high mean energy from an inertial perspective: The Unruh--Hawking effect acts like an amplification channel. We prove that a field in a Fock inertial state, probed via photon counting  by a non-inertial detector, realizes the optimal strategy attaining the ultimate sensitivity allowed by quantum mechanics for  the observation of the effect. We define the parameter regime in which the effect can be reliably revealed in laboratory experiments, regardless of the specific implementation.
\end{abstract}
\pacs{04.70.Dy, 03.65.Ta, 04.62.+v, 42.50.Dv}

\maketitle
%\begin{enumerate}

\noindent {\it Introduction.}---
The Unruh effect \cite{unruh_gravity_1976} is one of the most fundamental manifestations of the fact that the particle content of a  field theory is observer dependent \cite{unruhreview}.  A quantum (scalar) field in the Minkowski--Unruh vacuum from an inertial perspective, is detected as thermal by a Rindler observer in uniform acceleration. This is deeply connected with the phenomenon of Hawking radiation  \cite{hawking_gravity_1974}. In presence of an eternal black hole, if a scalar field is in the Hartle--Hawking vacuum as observed by a Kruskal observer (freely falling into the black hole), a Schwarzschild observer outside the event horizon would  detect, again, a thermal state \cite{birelli}. The temperature $T$ of the Unruh (Hawking)  thermal bath depends on the observer's acceleration $a$ (the black hole mass $M$ \cite{noteedu}):
\begin{equation}\label{temperature}
T_{\rm Unruh}=\frac{\hbar a}{2\pi c k_B }\,,\quad T_{\rm Hawking}=\frac{\hbar c^3}{8\pi G M k_B}\,,
\end{equation}
 where $c$ is the speed of light, $k_B$ is the Boltzmann constant, and $G$ is the gravitational constant.
 Despite their crucial role in modern theoretical physics, no experimental verification of the  Unruh--Hawking (UH) effects has been accomplished so far, as for conventional accelerations and in cosmological settings  the associated temperatures lie far below any  observable threshold. There have been different proposals to detect the Unruh effect via accelerated detectors \cite{unruhreview}, involving electrons in Penning traps, atoms in microwave cavities, backreaction  in ultraintense lasers, or Bose-Einstein condensates \cite{unruhdet}. Another path pursued in recent research is the detection of Hawking-like effects in  `artificial' black holes realized in liquids, optical fibers, or condensed matter systems  \cite{buchineri}, where a horizon analog  (e.g.~acoustic or optical)  may be created under controllable conditions.

In this Letter, borrowing rigorous methods from quantum statistics and estimation \cite{helstrom_estimation_1976}, we investigate the ultimate precision limits for the estimation of the UH temperature,  and we determine on fully general grounds the optimal conditions  for the revelation of UH effects. We are guided by an interesting  link between  field theory and quantum information: The change of coordinates between an inertial (or freely falling into a black hole) observer -- hereby called Alice -- and a non-inertial (or escaping the fall outside the horizon) observer -- hereby called Rob -- in the description of the state of a scalar field, is equivalent to the transformation that affects a light beam undergoing parametric down-conversion in an optical parametric oscillator   (see \cite{birelli,milburn_optics_1994}). Detection by an accelerated observer formally amounts to the action of a bosonic amplification channel.   We consider the realistic possibility that, in an experimental setup for the demonstration of UH effects, the field mode can be prepared in an arbitrary state from Alice's perspective, beyond the typical Minkowski--Unruh or Hartle--Hawking vacuum. We then devise the most suitable field states and the best detection schemes to be performed by Rob, in order to achieve optimal visibility and sensitivity in measuring the UH temperature. We show that having an increasingly large field energy from Alice's perspective results in a magnification of the UH signatures. We then prove that,  engineering the field in a Fock inertial state followed by Rob's photon  counting  (phonon counting in acoustic black hole analogs) allows for the optimal estimation of the UH effect. Alternative valid strategies involve coherent inertial states and Rob's heterodyne detections. We pin down the practical resources needed for an accessible and precise detection of the UH temperature. Our  findings are independent of the specialized implementation, setting a general goal for any experiment striving towards the unambiguous observation of the UH effects.

\medskip

\noindent {\it The Unruh effect as a bosonic amplification channel.}---
We consider a scalar field which is, from an inertial perspective, in a special  superposition of Minkowski monochromatic modes (see \cite{edu,ivetteinprep} for details) such that, in the Unruh basis \cite{birelli,edu,takagi}, Alice detects the field in the   single-mode state $\ket{\psi_0}_\omega$.
The annihilation operator of the mode satisfies the bosonic commutation relations: $[\hat{a}_{\omega}, \hat{a}_{\omega'}^\dagger]=\delta_{\omega,\omega'}$.
In Rindler coordinates the field is described as an entangled state of two  modes, truly monochromatic with  frequency $\omega$ \cite{edu} (from now on we drop the frequency subscript), each living in one of the two Rindler wedges, I (right) and II (left).
The Rindler field mode operators $\hat{b}_{\rm I,II}$ are connected to the Minkowski--Unruh ones via a Bogoliubov transformation \cite{birelli},
$\hat{a}=\cosh r\ \hat{b}_{\rm I} - \sinh r\ \hat{b}_{\rm II}$, where the `acceleration parameter' $r$ is proportional to the Unruh temperature:
$\cosh^{-2} r = 1-\exp(-\hbar \omega/k_B T)$.
 A non-inertial observer (Rob) in uniform acceleration $a$ is confined to Rindler region I.
 Thus the equilibrium state from Rob's viewpoint, in Schr\"odinger picture,
 is obtained by tracing over the modes in the causally disconnected region II:
 $\hat{\varrho}_r = {\rm Tr}_{\rm II} [\hat{U}(r) ({\ro_0} \otimes \ket{0}\!\bra{0}) \hat{U}^\dagger(r)]$.
Here $\hat{U}(r) = \exp[r (\hat b_{\rm I}^\dagger \hat b_{\rm II}^\dagger - \hat b_{\rm I} \hat b_{\rm II})]$ is the two-mode squeezing operator  that encodes the particle pair production between the two Rindler wedges (or across an eternal black hole horizon), and $\ro_0 \equiv \ket{\psi_0}\!\bra{\psi_0}$.
Such a phenomenon has a well known analog in quantum optics \cite{milburn_optics_1994}, which plays a crucial role for continuous variable quantum information \cite{gaussreview}.  An input signal beam in the state $\ro_0$ interacts with an idler mode (environment) in the vacuum via a two-mode squeezing transformation (realized by parametric down-conversion) with squeezing $r$. Tracing over the output idler mode, the output signal is left precisely in the mixed state $\ro_r$. Overall the non-unitary transformation from input to output, or from inertial to non-inertial frame, corresponds to the action of a bosonic amplification channel (see also \cite{bradler}) and can be described by the master equation ${d\ro_r/dr}=\tanh{r}\mathcal{L}[\hat{b}_{\rm I}^\dag]\ro_r$, where $\mathcal{L}[\hat{b}_{\rm I}^\dag]\ro_r=2\hat{b}_{\rm I}^\dag\ro_r \hat{b}_{\rm I}-\hat{b}_{\rm I}\hat{b}_{\rm I}^\dag\ro_r-\ro_r \hat{b}_{\rm I}\hat{b}_{\rm I}^\dag$. The solution to the master equation, using the disentanglement theorem \cite{milburn_optics_1994}, can be written as
 \begin{equation}\label{evoluto}
 \ro_r=N_r\sum_{k=0}^{\infty}C^k_r(\hat{b}_{\rm I}^\dag)^k(\cosh{r})^{-\hat{b}_{\rm I}^\dag \hat{b}_{\rm I}}\ro_0(\cosh{r})^{-\hat{b}_{\rm I}\hat{b}_{\rm I}^\dag}\hat{b}_{\rm I}^{k},\end{equation}
 with $N_r=\cosh^{-2}{r}$  and $C^k_r = (\tanh{r})^{2k}/{k!}$.
 \eq{evoluto}  precisely denotes the state detected by Rob who is non-inertial with acceleration parameter $r$, corresponding to a field mode state $\ro_0$, with mean photon number (energy), $\bar n_0={\rm Tr}[\ro_0 b_{\rm I}^\dagger b_{\rm I}]$ from Alice's inertial perspective.

 %For instance, if $\ro_0$ were the Minkowski--Unruh vacuum, then $\ro_r$ would be a thermal state with temperature $T$ [Eq.~(\ref{temperature})], proportional to $r$. Rob could then probe the field mode (of frequency $\omega$)   to extract informations about $r$, or $T$, hence to reveal the Unruh effect. In the following, we demonstrate that an arbitrarily better sensitivity in Rob's estimation of $r$ can be achieved if the field is in a state of non-minimal energy (mean photon number)  $\bar n_0={\rm Tr}[\ro_0 b_{\rm I}^\dagger b_{\rm I}]$ from Alice's perspective.

\medskip

\noindent {\it Optimal estimation of the UH effect}---
Suppose the following experiment is repeated $N$ times: the field is prepared in the state $\ro_0$ in the inertial frame, then a positive-operator-valued-measurement (POVM) $\{\hat{O}_\chi\}$ is performed by Rob on the modes in Rindler region I. Here  $\bra{\phi} \hat{O}_\chi\ket{\phi}\ge 0 \quad \forall\ket{\phi}$ and $\sum_\chi  \hat{O}_\chi=\mathbbm{1}$. For each strategy ${\cal S}=(\ro_0,  \hat{O})$, one can construct an unbiased estimator $\check{r}$ for $r$, of minimum variance \cite{cramer_estimation_1946} given by $N \mathrm{Var}[\check{r}] = I^{-1}_r({\cal S})$. Here the Fisher information $I_r({\cal S}) = \int d\chi p(\chi|r)\left(\frac{\partial\ln p(\chi|r)}{\partial r}\right)^2$, with $p(\chi|\theta)={\rm Tr}[{\hat{O}_\chi \ro_r}]$, is a figure of merit characterizing the performance of the strategy: the highest the Fisher information (FI), the most precise the estimation.  At a fixed $\ro_0$, the quantum Cram\'{e}r-Rao bound \cite{braunstein_quantum_1994} states that for any strategy it is $I_r({\cal S}) \le I(\ro_0,\hat{O}_{\hat{\Lambda}}) \equiv H_r(\ro_0)$, i.e., there exists an optimal POVM yielding maximum sensitivity, that consists of projections on the eigenstates of the so-called `symmetric logarithmic derivative' $\hat{\Lambda}_{\ro_r}$, an observable which depends on $\ro_r$ and is defined implicitly as follows,
$2 {d \ro_r}/{d r} =  \hat{\Lambda}_{\ro_r} \ro_r + \ro_r \hat{\Lambda}_{\ro_r}$. The FI associated to such optimal measurement is known as quantum FI (QFI), $H_r(\ro_0) = {\rm Tr} [\ro_r \hat{\Lambda}_{\ro_r}^2]$.
%\textbf{In the following we will consider we work with a large $N$ such that
%we are allow to realize one-step adaptive scheme, that is to use a vanishing
%number of copies to make a rough estimate $\hat{r}_0$ and then measure
%$\Lambda(\hat{r_0})$ on the outstanding copies to polish up the estimate $\hat{r}$}.
Alternatively,  the QFI can be computed from the Bures metric \cite{wootters}, which in turn is related to the quantum fidelity \cite{jozsa_quantum_1994} $\mathcal{F}(\hat{\varrho}_1,\hat{\varrho}_2)= (\tr [\sqrt{\sqrt{\hat{\varrho}_1}\hat{\varrho}_2\sqrt{\hat{\varrho}_1}} ] )^2$ between two infinitesimally close states (in our case, evolved from the same $\ro_0$):
$H_r(\ro_0) = 4 [1-{\cal F}(\ro_{r},\ro_{r+dr})]/dr^2$. We will now investigate strategies for the estimation of $r$, involving Gaussian (coherent, squeezed) or non-Gaussian (Fock) field states from Alice's perspective beyond the typical Minkowski--Unruh vacuum, and we will aim for those with the highest possible (quantum) Fisher information \cite{notemonras,monras_quantum_2007,adesso_quantum_2008,monras_channels_2010}.

%We will consider two classes of states, pure Gaussian (displaced squeezed) states, and Fock states. The latter states are shown to be optimal for this estimation task, attaining the ultimate bound as in the case of a damping channel.

\medskip

\begin{figure}[t]
\includegraphics[width=8cm]{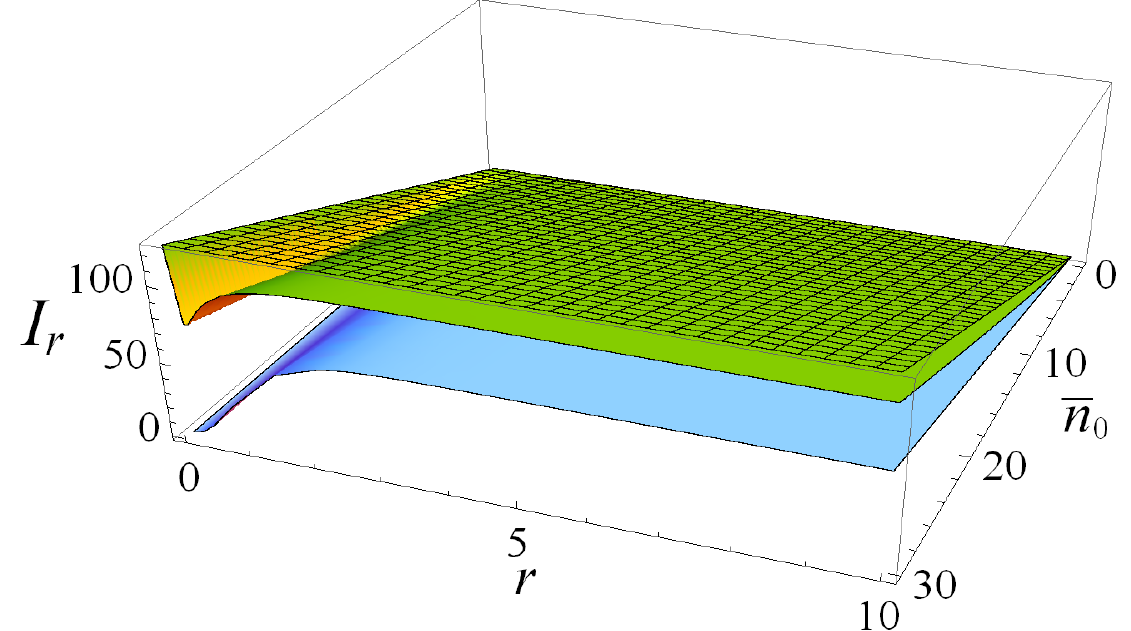}
\caption{(Color online) Fisher informations $I_r$ corresponding, from bottom to top, to: coherent states in Alice's frame and heterodyne detection by Rob (blue); general Gaussian states in Alice's frame and optimal detection by Rob (green); ultimate quantum bound, attained by Fock states in Alice's frame and photon counting by Rob (wireframe).
The variance in estimating  $r$ scales as $1/I_r$.}
\label{fishfingers}
\end{figure}

\noindent{\it Gaussian field states.}--- Gaussian states, e.g.~ground and thermal states of harmonic oscillators, play an important role in quantum optics and many-body physics since they are easy to manipulate mathematically and provide a good description of states commonly produced in experiments. We begin by considering a displaced squeezed pure Gaussian field state $\ro^G_0$ in the inertial frame. It can be completely specified by its first moments, $\xi_0=(q_0, p_0)^T$, and its covariance matrix \cite{gaussreview}
\[
\gr\sigma_0 = \left(
\begin{array}{cc}
 e^{2 s} \cos ^2 \theta  +e^{-2 s} \sin ^2 \theta   & \sin 2 \theta
   \sinh  2 s  \\
 \sin  2 \theta  \sinh  2 s  & e^{-2 s} \cos ^2 \theta +e^{2 s} \sin
   ^2 \theta
\end{array}
\right)\!.
\]
Here $s>0$ is the squeezing degree, with phase  $\theta$.
Under the action of the bosonic amplification channel associated with the UH effect, \eq{evoluto}, the Gaussian state is transformed into another Gaussian state $\ro_r$ characterized by the following first and second moments \cite{wolfevol}:  $\xi_r = X_r \xi_0$, $\gr\sigma_r = X_r \sigma^{in} X_r^T + Y_r$, where $X_r = \cosh r\ \mathbbm{1}_2$ and
$Y_r=\sinh^2 r\ \mathbbm{1}_2$.
In order to compute the QFI via the Bures metric \cite{calsami} (see also \cite{monras_channels_2010}), we recall that the fidelity between two arbitrary one-mode Gaussian states $\ro^G_1$ and $\ro^G_2$ with respective covariance matrix $\gr\sigma_{1,2}$ and first moments $\xi_{1,2}$, is ${\cal F}(\ro^G_1,\rho^G_2) = 2(\sqrt{\Sigma+\Gamma}-\sqrt{\Gamma})^{-1}\exp [-(\xi_2-\xi_1)^T(\gr\sigma_1+\gr\sigma_2)^{-1}(\xi_2-\xi_1)]$ \cite{scutaru}, where $\Sigma = \det(\gr\sigma_1+\gr\sigma_2)$, and $\Gamma = (\det \gr\sigma_1-1)(\det\gr\sigma_2-1)$. In our case, we note that the fidelity ${\cal F}\left(\ro^G_r,\ro^G_{r+dr}\right)$ is minimized for $\theta=0$ and $q_0=0$. The inertial energy of the Gaussian field  becomes
$\bar n_0 = \sinh^2 s + p_0^2/2$. By setting $\sinh^2 s = x \bar n_0$ and $p_0^2 =2 (1-x) \bar n_0$, we introduce an energy ratio $x$, ranging from $0$ (purely coherent inertial state) to $1$ (squeezed inertial state) \cite{monras_quantum_2007}. We finally obtain the QFI:
$H_r [\ro^G_0(\bar n_0,x)] =\{[2 (x-1) \bar{n}_0]/[\cosh (2 r) (x \bar{n}_0+1)-\sqrt{(x
   \bar{n}_0)(x \bar{n}_0+1)}]\}  +  \{16 x \bar{n}_0/[\cosh (4 r) (x
   \bar{n}_0+1)-x \bar{n}_0+3]\}-2 (x-1) \bar{n}_0 -2 (x-1) \bar{n}_0
   \sqrt{x \bar{n}_0/(x \bar{n}_0+1)}+4$. The maximal QFI is obtained by numerical optimization  over $x$ for any given $\bar n_0$ and $r$.
   For very small $r$, squeezed states from Alice's perspective are optimal ($x=1$), while for higher $r$ a nonzero displacement improves the estimation.
   %For $r \rightarrow \infty$, the optimal $x$ converges to $x=\bar n_0 / [2+3 \bar n_0+2 (1+\bar n_0)^{3/2}]$.
   The maximal $H_r(\ro^G_0)$ is a monotonically increasing function of the energy $\bar n_0$ measured in the inertial frame (See Fig.~\ref{fishfingers}, middle surface). Therefore using suitable engineered Gaussian field modes one can significantly improve the sensitivity in the detection of the UH effect compared to the case of a Minkowski--Unruh vacuum, whose variance $H^{-1}_r[\ro^G_0(0,0)]=1/4$ is actually the largest.  However,   this strategy may be hard to implement in practice, as the optimal Rindler measurement depends on  $r$ thus requiring an adaptive estimation scheme \cite{helstrom_estimation_1976,monras_quantum_2007}.

We can then consider the following simpler strategy. Let $\ro^G_0 \equiv \ket {\alpha_0} \! \bra {\alpha_0}$ be just a coherent state (with $\alpha_0 = (q_0 + i p_0)/\sqrt2$). Rob detects the field in a displaced thermal state with energy $\vert{\alpha_r}\vert^2=\cosh^2{r}\vert{\alpha_0}\vert^2$.  Let the POVM $\{\hat{O}_\chi\}$  be a heterodyne detection, i.e. projection  onto a set of coherent states, $\{1/\pi\ket{\chi}\bra{\chi}\}$. The FI associated to this strategy is $I_r(\ket{\alpha_0},\ket{\chi}\!\bra{\chi})=4(1+\bar{n}_0/2)\tanh^2{r} < H_r[\ro^G_0(\bar{n}_0,0)]$. Although suboptimal, this strategy still encodes an increase in sensitivity with increasing inertial energy (displacement), as shown in  Fig.~\ref{fishfingers} (bottommost surface). Namely, if $\bar{n}_0 \equiv |\alpha_0|^2 \gg 2/\sinh^2 r$, then the coherent strategy combines practical feasibility with an arbitrarily good improvement over the conventional case of a Minkowski--Unruh vacuum.

\medskip

\noindent {\it Fock field states.}---
We turn now to explore an estimation strategy involving non-Gaussian field states.
Let the state be a Fock state $\ro^F_0 = \ket{n_0}\!\bra{n_0}$ from Alice's perspective, for which trivially $\bar n_0 = n_0$.
From \eq{evoluto}, the state as detected by Rob is  $\ro^F_r = \sum_{k=0}^{\infty} c_{n,k}(r) \ket{n_0+k}\!\bra{n_0+k}$, with
$c_{n,k}(r) = \binom{n+k}{n}(\cosh r) ^{-2 (n+1)} \tanh^{2k} r$.
In this case, to find the optimal measurement strategy, it is more convenient to use the definition of the QFI in terms of the symmetric logarithmic derivative, which in turn can be easily computed once the spectrum of $\ro^F_r$ is known. We find
\begin{equation}\label{hfock}
H_r(\ket{\bar n_0}) = \sum_{k=0}^\infty \frac{[\partial c_{\bar n_0,k}(r)/\partial r]^2}{c_{\bar n_0,k}(r)} =4(1+\bar n_0)\,.  \end{equation}
As shown in Fig.~\ref{fishfingers} (topmost surface), the QFI for Fock inertial states beats the optimal Gaussian one in the whole parameter space (they only coincide in the limit $r \rightarrow 0$), allowing for a significantly reduced variance in the estimation of $r$ at fixed $\bar n_0$.
The optimal measurement strategy is in this case simply photon counting (independently of the value of $r$). Thus by having a field which is prepared  in a Fock state with high enough energy from an inertial perspective (in the Unruh basis), one can estimate the UH acceleration parameter -- i.e., reveal the effect -- with {\em arbitrarily high sensitivity} by simply letting Rob count photons in the field he detects.

\medskip

\noindent{\it Ultimate quantum bound.}--- We will now prove that no improvement over the above Fock--based quantum estimation strategy can be achieved  even allowing causality violation. Suppose a hypothetical  observer  existed, able to measure jointly the field in the two spacelike separated Rindler regions, i.e., able to estimate globally the two-mode (unitary) squeezing transformation of coordinates $\hat{U}(r)$, so as to extract an ultimately precise estimator for $r$ without any information loss. The QFI $H^{\max}_r$ in this unphysical setup  can be written as $H_r^{\max}=4(\avr{\hat G^2}-\avr{\hat G}^2)$ \cite{helstrom_estimation_1976}, where  $\hat G = -i (\hat b_{\rm I}^\dagger \hat b_{\rm II}^\dagger - \hat b_{\rm I} \hat b_{\rm II})$ and the average is taken over the state of system plus idler (from an inertial perspective), $\ro_0 \otimes \ket{0}\!\bra{0}$.  We have $\avr{\hat G}=0$ and
$\avr{\hat G^2} = 1+\bar n_0$. Hence the ultimate quantum bound on the estimation of the UH effect is given by $H_r^{\max} = 4(1+\bar n_0)$, exactly equal to the QFI in Eq.~(\ref{hfock}). We conclude that, most notably, Fock field states in the inertial frame, in conjunction with photon counting performed by Rob (physically confined to  Rindler region I), allow for the absolutely optimal estimation of $r$: {\em no advantage} could be achieved even if access to the degrees of freedom in Rindler region II was permitted.
%This  is in analogy to the case of a dissipation channel \cite{adesso_quantum_2008}.

\medskip

\noindent {\it Discussion.}--- To draw more practical conclusions, it is convenient to derive figures of merit associated with the direct estimation of the UH thermal (amplification) energy $n_T=\sinh^2 r$, which amounts to the mean photon number of a thermal mode with temperature $T$ [\eq{temperature}]. Using \eq{hfock} and the transformation rule for the FIs  \cite{helstrom_estimation_1976},
 $I_{n_T} = I_r (\partial r / \partial n_T)^2$, the minimum variance corresponding to the optimal detection of $n_T$ becomes:
 $N \mathrm{Var}^{\min}[\check{n}_T] = (n_T+n_T^2)/(1+\bar n_0)$.
 %The variance is proportional to the UH energy itself $n_T$, that is the quantity one wants to reveal, but once again we note that it can be arbitrarily reduced by increasing the ``in'' energy $\bar n_0$.
 For a single experimental run ($N=1$), the relative error on the estimation of $n_T$ is defined as $\varepsilon_{n_T}=(\mathrm{Var}^{\min}[\check{n}_T])^{1/2}/n_T = [(1+n_T^{-1})/(1+\bar n_0)]^{1/2}$. For a  precise estimation, it must be  $\varepsilon_{n_T} \ll 1$, i.e., \begin{equation}\bar n_0  \gg n_T^{-1}.\label{threshold}\end{equation} This result defines the regime for optimal sensitivity \cite{notefuck}, and we will now link it with a simple assessment of visibility of the UH effect.  From \eq{evoluto} we have, for a generic field state with mean energy $\bar n_0$ as detected by Alice, that the mean energy as detected by Rob is  $\bar n_r={\rm Tr}[\ro_r \hat{b}_{\rm I}^\dagger \hat{b}_{\rm I}] = \bar n_0 + n_{T} (\bar n_0 +1)$. To make the UH effect observable, the energy difference (visibility) $\Delta \bar n = \bar n_r - \bar n_0$ must be at least of the order of one photon: $\Delta \bar n \gtrsim \max\{1,n_T\}$. Interestingly, the threshold in precision, \eq{threshold}, translates in a visibility $\Delta \bar n \gg 1+n_T$, i.e., precisely the regime that renders the effect amenable of detection. We  conclude that  {\em \eq{threshold} sets the ideal threshold on the inertial field energy for any reliable -- both accessible and accurate -- verification of the UH effect}.

%   This is important e.g.~for all those proposals that aim at detecting the Unruh effect by accelerating $k$-level particles, which play the role of the ``out'' observers, and then measuring the excitation/deexcitation rate as a consequence of the interaction with the UH bath \cite{unruhreview}. Such a rate is magnified and more easily measurable, at a given acceleration of the particles, when \eq{threshold} holds.

\begin{figure}[t]
\includegraphics[width=7.5cm]{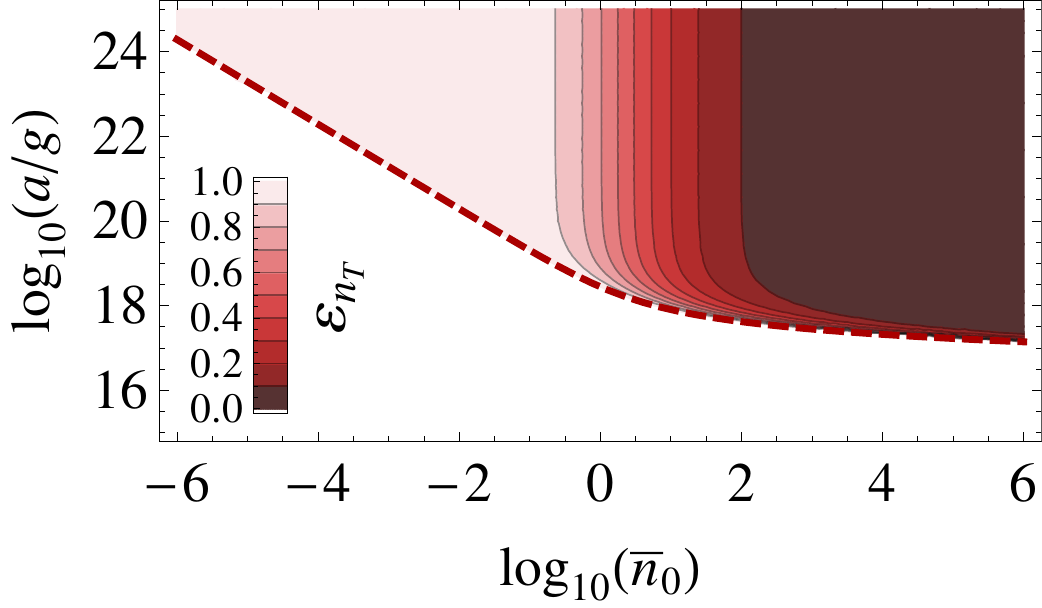}
\caption{(Color online) Contour plot of the relative error $\varepsilon_{n_T}$ on the estimation of the Unruh thermal energy    versus the acceleration $a$ of Rindler detectors and the energy $\bar n_0$ of a microwave field in an inertial frame (in log scales). The thick dashed line, \eq{threshold}, sets the threshold for  accessible and precise detection of the Unruh effect.}
\label{contornino}
\end{figure}

Additional fine-tuning is certainly needed  to adapt our general prescription to the facilities of specific  proposals to measure the Unruh effect or to mimic Hawking radiation \cite{unruhdet,buchineri}; this lies beyond the scope of this Letter. Just to have a flavour of the involved orders of magnitude, let us consider  the setting of Rindler detectors (e.g., two-level atoms) accelerated in microwave fields ($\omega \approx 10^{10} {\rm Hz}$) \cite{unruhreview}. For this example, we plot in Fig.~\ref{contornino} the relative error $\varepsilon_{n_T}$ in the estimation of $n_T$,
 versus the acceleration $a$ of the detector and the microwave field energy $\bar n_0$ in the inertial frame.
 While for $\bar n_0=10^{-6}$ (a quasi-vacuum field) an acceleration of at least $10^{25} g$ is needed to detect the Unruh bath, already
 with $\bar n_0 =1$ photon the threshold (determined by equality in \eq{threshold}) drops to a more accessible $10^{18} g$: a dramatic magnification of the Unruh effect.
% The estimation error $\varepsilon_{n_T}$ becomes vanishingly small when $\bar n_0$ is at least of the order of $10^2-10^3$ photons.
 The error is further reduced by a statistical factor $\sqrt{N}$ by repeating the experiment $N$ times.
%, resulting in a further decrease of the acceleration required to reveal the UH effect.
These are promising findings in view of the recent progresses in the production of Fock states in microwave cavities and circuit QED \cite{fock}, and in the degree of  control of the photon counting technique \cite{counting}.

\medskip

\noindent{\it Conclusions.}--- We have proven that the UH effects are magnified when a scalar field is in a state of non-minimal mean energy, e.g.~coherent or Fock, from an inertial perspective. Accessible and precise measurement of the UH temperature is enabled by heterodyne detection or photon counting performed by a non-inertial observer. Beyond a fundamental interest, our findings are of direct practical relevance, delivering clearcut prescriptions for the optimal revelation of the UH effects, independent of the specific implementation.
The techniques developed here might be applied to other field theory phenomena such as the dynamical Casimir effect.

\smallskip

\noindent{\it Acknowledgments.}--- We thank J. Calsamiglia, H. Culetu, T. Downes, S. Fagnocchi, F. Illuminati, P. Kr\"uger, J. Louko, E. Mart\'in-Mart\'inez, A. Monras, R. Mu\~{n}oz-Tapia, M. Paternostro,  and M. M. Wolf for discussions. We were supported by EPSRC [CAF Grant EP/G00496X/2], Spanish MICINN [Contract FIS2008-01236] and CONSOLIDER 2006-2010~[CIRT contract 2009SGR-0985].

%\end{enumerate}


\begin{thebibliography}{99}



\bibitem{unruh_gravity_1976}
 W.~G. Unruh, Phys. Rev. D {\bf 14},
870,
(1976).

\bibitem{unruhreview} L. Crispino {\it et al.}, Rev. Mod. Phys. {\bf 80}, 787 (2008).

\bibitem{hawking_gravity_1974}
 S.~W. Hawking, Nature {\bf 248},
30  (1974); Commun. Math. Phys. {\bf 43},
199 (1975).

\bibitem{birelli} N. D. Birrell and P. C. W. Davies,
{\it Quantum fields in curved space} (Cambridge University Press,
Cambridge, 1982).

\bibitem{noteedu} In general $T$ has an inverse dependence on the observer's distance $d$ from the horizon, that is relevant for small $d$ \cite{edu}.

\bibitem{edu} E. Mart\'in-Mart\'inez {\it et al.}, Phys. Rev. D {\bf 82}, 064006 (2010).


\bibitem{unruhdet}
J. Rogers, Phys. Rev. Lett. {\bf 61}, 2113  (1988); M. L. Scully {\it et al.},  {\it ibid.} {\bf 91}, 243004 (2003);
P. Chen and T. Toshi,  Phys. Rev. Lett. {\bf 83}, 256 (1999); R.  Sch\"utzhold, G. Schaller, and D. Habs,  {\it ibid.} {\bf 97}, 121302 (2006); A. Retzker {\it et al.}, {\it ibid.} {\bf 101}, 110402 (2008).

\bibitem{buchineri}  W.~G. Unruh, Phys. Rev. Lett. {\bf 46},
 1351 (1981);  L. J. Garay {\it et al.}, Phys. Rev. Lett. {\bf 85}, 4643 (2000); U. Leonhardt,  T. Kiss, and P. Ohberg,  J. Opt. B {\bf 5}, S42 (2003); T. G. Philbin {\it et al.},  Science {\bf 319}, 1367 (2008); P. D. Nation {\it et al.}, Phys. Rev. Lett.  {\bf 103}, 087004 (2009); I. Carusotto {\it et al.}, New J. Phys. {\bf 10}, 103001  (2008).

\bibitem{helstrom_estimation_1976} C. W. Helstrom, {\it Quantum detection and estimation theory} (Academic Press, NY, 1976); A. S. Holevo,  {\it Probabilistic and statistical aspects of quantum theory} (North-Holland, Amsterdam, 1982).



\bibitem{milburn_optics_1994}
D.~F. Walls and G.~J. Milburn, \emph{Quantum Optics}\ (Springer,
  Berlin, 1995).



%\bibitem{singlemode} Our setting is completely general and does not rely on the so-called single-mode approximation, which was instrumental to pioneering studies in relativistic quantum information but has recently proven to be incorrect \cite{edu,ivetteinprep}.


\bibitem{ivetteinprep} D. Bruschi {\it et al.}, arXiv:1007.4670.

%\bibitem{noteheis}Notice that in this picture states are evolved at fixed mode operators, hence in what follows $\hat{a}$ is identified with $\hat {b}_I$.


\bibitem{takagi} S. Takagi, Prog. Theor. Phys. Suppl. {\bf 88}, 1 (1986).

\bibitem{gaussreview} G. Adesso and F. Illuminati, J. Phys. A {\bf 40}, 7821 (2007).

\bibitem{bradler} K. Br\'adler {\it et al.}, JHEP {\bf 0908}, 074 (2009).




\bibitem{cramer_estimation_1946} H. Cram\'{e}r, {\it Mathematical Methods of Statistics} (Princeton University Press, 1946).

\bibitem{braunstein_quantum_1994} S.L. Braunstein and C. M. Caves, Phys. Rev. Lett. {\bf 72}, 3439 (1994).

\bibitem{wootters} W. K. Wootters, Phys. Rev. D {\bf 23}, 357 (1981).

\bibitem{jozsa_quantum_1994} R. Jozsa, J. Mod. Opt. {\bf 41}, 2315 (1994).


\bibitem{notemonras} These techniques have been recently applied to  the optimal estimation of lossy channels \cite{monras_quantum_2007,adesso_quantum_2008}, and  general  Gaussian channels probed by Gaussian states \cite{monras_channels_2010}.
    %The bosonic amplification channel is an instance of the latter, but we also allow for non-Gaussian inertial probes,  not covered by the framework of Ref.~\cite{monras_channels_2010}.

\bibitem{monras_quantum_2007} A.\,Monras\,and\,M.G.A.\,Paris,\,Phys.\,Rev.\,Lett.\,{\bf 98},\,160401\,(2007).

\bibitem{adesso_quantum_2008} G. Adesso {\it et al.}, Phys. Rev. A. {\bf 79}, 040305(R) (2009).


\bibitem{monras_channels_2010} A. Monras and F. Illuminati, Phys. Rev. A {\bf 81}, 062326 (2010).

\bibitem{wolfevol} A. Serafini {\it et al.}, J. Opt. B {\bf 7}, R19 (2005).

\bibitem{calsami} J. Calsamiglia {\it et al.}, Phys. Rev. A.  {\bf 77}, 032311 (2008).

\bibitem{scutaru} H. Scutaru, J. Phys. A {\bf 31}, 3659 (1998).

\bibitem{notefuck} This holds for the Fock strategy. For coherent  fields, the RHS of \eq{threshold} must be multiplied by a factor 2.


\bibitem{fock} B. T. H. Varcoe {\it et al.}, Nature {\bf 403}, 743 (2000);
P. Bertet \emph{et al.}, Phys. Rev. Lett. {\bf 88}, 143601 (2002);
J. McKeever \emph{et al.}, Science {\bf 303}, 1992 (2004);  A. A. Houck {\it et al.}, Nature {\bf 449}, 328 (2007);
S. Deleglise \emph{et al.}, Nature {\bf 455}, 510 (2008);
M. Hofheinz {\it et al.}, Nature {\bf 454}, 310 (2008).

\bibitem{counting} D. I. Schuster {\it et al.}, Nature {\bf 445}, 515 (2007); C. Guerlin {\it et al.}, Nature {\bf 448}, 889 (2007);
A. Divochiy {\it et al.}, Nat. Photon. {\bf 2}, 302 (2008).


%\bibitem{photonote} Imperfect photon counting  induces a correction of \eq{threshold} as $\bar n_0>(\tau n_T)^{-1}$, $\tau \le 1$ being the photodetection efficiency.











\end{thebibliography}
\end{document}